\documentclass[superscriptaddress,aps,pra,twocolumn,showpacs,nofootinbib,longbibliography]{revtex4-1}
\usepackage{etex}
\usepackage{amsmath,amssymb,amsthm}
\usepackage[colorlinks=true,citecolor=blue,urlcolor=blue]{hyperref}
\usepackage[pdftex]{graphicx}
\usepackage{times,txfonts}
\usepackage{braket}
\usepackage{color}
\usepackage{natbib}

\newcommand{\be}{\begin{equation}}
\newcommand{\ee}{\end{equation}}
\newcommand{\ba}{\begin{eqnarray}}
\newcommand{\ea}{\end{eqnarray}}

\begin{document}

\title{Sharing of tripartite nonlocality by multiple observers measuring sequentially at one side}

\author{Sutapa Saha}
\email{sutapa.gate@gmail.com}
\affiliation{Physics and Applied Mathematics Unit, Indian Statistical Institute, 203 B. T. Road, Kolkata 700 108, India.}

\author{Debarshi Das}
\email{dasdebarshi90@gmail.com}
\affiliation{Centre for Astroparticle Physics and Space Science (CAPSS), Bose Institute, Block EN, Sector V, Salt Lake, Kolkata 700 091, India}

\author{Souradeep Sasmal}
\email{souradeep.007@gmail.com}
\affiliation{Centre for Astroparticle Physics and Space Science (CAPSS), Bose Institute, Block EN, Sector V, Salt Lake, Kolkata 700 091, India}

\author{Debasis Sarkar}
\email{dsappmath@caluniv.ac.in}
\affiliation{Department of Applied Mathematics, University of Calcutta, 92, A.P.C. Road, Kolkata-700 009, India}

\author{Kaushiki Mukherjee}
\email{kaushiki_mukherjee@rediffmail.com}
\affiliation{Department of Mathematics, Government Girls' General Degree College, Ekbalpore, Kolkata--700 023, India.}

\author{Arup Roy}
\email{arup145.roy@gmail.com}
\affiliation{Physics and Applied Mathematics Unit, Indian Statistical Institute, 203 B. T. Road, Kolkata 700 108, India.}

\author{Some Sankar Bhattacharya}
\email{somesankar@gmail.com}
\affiliation{Physics and Applied Mathematics Unit, Indian Statistical Institute, 203 B. T. Road, Kolkata 700 108, India.}

\begin{abstract}
Standard tripartite nonlocality and genuine tripartite nonlocality can be detected by the violations of Mermin inequality and Svetlichny inequality, respectively. Since tripartite quantum nonlocality has novel applications in quantum information and quantum computation, it is important to investigate whether more than three observers can share tripartite nonlocality, simultaneously. In the present study we answer this question in the affirmative. In particular, we consider a scenario where three spin-$\frac{1}{2}$ particles are spatially separated and shared between Alice, Bob and multiple Charlies. Alice performs measurements on the first particle; Bob performs measurements on the second particle and multiple Charlies perform measurements on the third particle sequentially. In this scenario we investigate how many Charlies can simultaneously demonstrate standard tripartite nonlocality and genuine tripartite nonlocality with single Alice and single Bob. The interesting result revealed by the present study is that at most six Charlies can simultaneously demonstrate standard tripartite nonlocality with single Alice and single Bob. On the other hand, at most two Charlies can simultaneously demonstrate genuine tripartite nonlocality with single Alice and single Bob. Hence, the present study shows that standard tripartite nonlocality can be simultaneously shared by larger number of Charlies compared to genuine tripartite nonlocality in the aforementioned scenario, which implies that standard tripartite nonlocality is more effective than genuine tripartite nonlocality in the context of simultaneous sharing by multiple observers.

\end{abstract}

\maketitle

\section{Introduction} \label{intro}

Nonlocality is one of the salient features of quantum mechanics. It was initially pointed out by Einstein, Podolsky and Rosen in their famous EPR paper \cite{epr} which challenged some of the greatest thinkers in physics and philosophy on the consistency of the notion of local-realism with quantum mechanics. It was John Bell who first rescued this debate from metaphysical argument to experimentally testable criteria, known as Bell's inequality \cite{Bell}.  Bell's inequality demonstrates that quantum mechanics possesses some correlations which cannot be described by local realistic theory or local hidden variable models \cite{Bell}. Bell’s inequality was later improved for real experimental purposes by Clauser, Horne, Shimony and Holt (CHSH) \cite{chsh}. Entanglement \cite{ent} in quantum systems is shown to be necessary for demonstrating quantum nonlocality. However, the converse may not be true always. Quantum nonlocal correlations are found to be fundamental resources in various information processing tasks, such as device independent randomness generation \cite{appran}, key distribution \cite{appkd1,appkd2}, reductions of communication complexity \cite{appcc} and many more. 

Since quantum nonlocal correlations are proved to be precious quantum information theoretic resources, it is legitimate to ask whether nonlocality can be shared between multiple observers. However, contrary to classical correlations, quantum correlations cannot be shared between arbitrary number of spatially separated systems, which is quantitatively expressed through the monogamy relations for 
entanglement \cite{monen} or the monogamy relations for nonlocality \cite{monnon}. One important point to be stressed here is that the no-signalling condition (the probability of obtaining one party's outcome does not depend on spatially separated other party's setting), a consequence of relativistic causality principle, is satisfied between any pair of observers in case of the multipartite scenario considered in the monogamy context. Hence, it is interesting to ask whether sharing of nonlocality among multiple observers can be improved invoking a scenario where no-signalling assumption is partially relaxed without violating relativistic causality. Silva et. al. addressed the above issue in the affirmative by considering a scenario that involves sequential measurements on one particle by multiple observers \cite{sygp}. In particular, the scenario is that one observer (say, Alice) in one wing and multiple observers (say, multiple Bobs) in the other wing share an entangled system of two spin-$\frac{1}{2}$ particles, where Alice is spatially separated from multiple Bobs. Alice performs measurements on one spin-$\frac{1}{2}$ particle and multiple Bobs measure on another spin-$\frac{1}{2}$ particle sequentially. In this scenario it has been shown  numerically \cite{sygp} as well as analytically \cite{mal} that at most two Bobs can simultaneously demonstrate nonlocality with single Alice with respect to the quantum violations of CHSH inequality, when each of the multiple Bobs performs different measurements with equal probability and when the measurements of each Bob are independent of the choices of measurement settings and outcomes of previous Bobs. Experiments have also been carried out confirming the above result \cite{exp1,exp2}. Note that monogamy relations for nonlocality states that two observer cannot simultaneously demonstrate nonlocality with respect to the quantum violations of CHSH inequality with a third observer \cite{monnon}. 

Recently, quantum steering \cite{steer1,steer2,steer3} of a single system sequentially by multiple observers in the above scenario has also been demonstrated \cite{sas}, going beyond monogamy constraint of quantum steering \cite{monsteer1,monsteer2}. Sharing of entanglement by multiple observers measuring sequentially in the above scenario has been demonstrated \cite{bera}. Sharing of nonlocal advantage of quantum coherence \cite{naqc} in the above scenario has also been presented recently \cite{datta}.

Quantum nonlocality is well understood in the bipartite scenario. However, multipartite nonlocality  is significantly less explored than its bipartite counterpart and the complexity increases in multipartite scenario resulting in more elegant and sophisticated foundational significances. Mermin derived Bell-type inequality in order to capture standard tripartite nonlocality (or, Bell nonlocality) \cite{mermin}. However, Mermin's inequality does not incorporate ``genuineness" of tripartite nonlocality.  A multipartite state is called genuine nonlocal if and only if the state is nonlocal with respect to any possible partitions of the multipartite state. For a tripartite system, the conventionally employed inequality, whose violation implies genuine tripartite nonlocality, is due to Svetlichny \cite{SI}. Mermin inequality and Svetlichny inequality can be violated by both Greenberger-Horne-Zeilinger (GHZ) \cite{GHZ} and W classes of states for three qubits \cite{vio1,vio2,vio3}. 

Besides having foundational significances \cite{foundsig}, multipartite quantum correlations are used as resources in quantum communication and quantum computation as evidenced by a wide range of theoretical and experimental studies \cite{ap1,ap2,ap3,ap4,ap5,ap6,ap7,ap8,ap9,ap10,ap11,ap12,ap13,ap14}. Hence, it is significant to investigate how $n$-partite quantum correlation  can be used as resource by more than $n$ observers, simultaneously, by performing sequential measurements. Surprisingly, all the previous studies addressing quantum correlation sharing by multiple sequential observers were restricted to two spatially separated particles. The generalization of the above feature of quantum correlation to multiple spatially separated particles remains unexplored till date. Motivated by these facts, in the present study we investigate sharing of tripartite nonlocality in the scenario where three spin-$\frac{1}{2}$ particles are spatially separated and shared between, say, Alice, Bob and multiple Charlies. Alice measures on the first particle; Bob measures on the second particle and multiple Charlies measure on the third particle sequentially. In this scenario we investigate how many Charlies can demonstrate standard tripartite nonlocality as well as genuine tripartite nonlocality through quantum violations of Mermin inequality \cite{mermin} and Svetlichny inequality \cite{SI}, respectively. Besides providing a new dimension in the context of nonlocality sharing, the present study seeks to demonstrate a new method to operationally distinguish two inequivalent forms of tripartite nonlocality, standard and genuine tripartite nonlocality.

The paper is organized as follows: in Sec. \ref{nonlocality} we recapitulate the notions of standard as well as genuine tripartite nonlocality followed by Sec. \ref{scenario}, where we describe in detail the measurement scenario considered in the present study and the unsharp measurement formalism. In Secs. \ref{sharetri} and \ref{sharetri2} we investigate sharing of standard tripartite nonlocality and genuine tripartite nonlocality, respectively, contingent upon using sequential measurements in the scenario considered. Finally, in Sec. \ref{conclusion} we present the concluding discussion.

\section{Recapitulating tripartite nonlocality} \label{nonlocality}

Now we are going to discuss the concept of tripartite nonlocality. Consider a tripartite Bell scenario where each of three spatially separated parties, say, Alice, Bob and Charlie performs two dichotomic measurements on their subsystems. In this scenario the correlation is described by the conditional probability distributions $P(a, b, c|A_x, B_y, C_z)$, where $x,y,z\in\{0,1\}$ and $a,b,c\in\{+1,-1\}$. $x$, $y$, $z$ denote the choices of measurement settings of Alice, Bob and Charlie, respectively; $a$, $b$, $c$ denote the outcomes of Alice, Bob and Charlie, respectively. The correlation $P(a, b, c|A_x, B_y, C_z)$ is fully local iff, for all $x$, $y$, $z$, $a$, $b$, $c$, it can be explained by a fully local hidden variable (LHV) model given by,
\begin{equation}
P(a, b, c|A_x, B_y, C_z)=\sum_{\xi} p(\xi) P(a|A_x, \xi)P(b|B_y, \xi)P(c|C_z, \xi), \label{FLHV}
\end{equation}
where $p(\xi)$ is the probability distribution over the hidden variables $\xi$; $0\leq p(\xi)\leq 1$ and $\sum_{\xi} p(\xi) = 1$. $P(a|A_x, \xi)$ denotes the conditional probability of getting outcome $a$ when Alice performs the measurement $A_x$ on her subsystem and $\xi$ is the hidden variable. $P(b|B_y, \xi)$, $P(c|C_z, \xi)$ are similarly defined. The correlation $P(a, b, c|A_x, B_y, C_z)$ exhibits standard tripartite nonlocality (i. e., Bell nonlocality) iff it is not fully local.

Standard tripartite nonlocality is detected by violation of  the Mermin inequality \cite{mermin}, which has the following form,
\begin{equation}
M = | \langle A_1 B_0 C_0 \rangle + \langle A_0 B_1 C_0 \rangle  + \langle A_0 B_0 C_1 \rangle - \langle A_1 B_1 C_1 \rangle | \leq 2.
\label{mi}
\end{equation}
Here $\langle A_xB_yC_z \rangle=\sum_{abc} (a  b c) P(a, b, c|A_x, B_y, C_z)$. The maximum violation of the Mermin inequality in quantum mechanics is $4$. Note that Mermin inequality is maximally violated by a GHZ state \cite{GHZ} and the measurement settings that give rise to it exhibit the GHZ paradox \cite{mermin2}. 

If a correlation violates Mermin inequality, it does not necessarily imply that the correlation exhibits genuine tripartite nonlocality \cite{SI,B}. Svetlichny introduced the strongest form of genuine tripartite nonlocality in Ref. \cite{SI} (see Ref. \cite{B} for the other two forms of genuine nonlocality).
A correlation demonstrates genuine tripartite nonlocality iff it does not have the following hybrid nonlocal-local hidden variable (NLHV) model,
\begin{align}
P(a, b, c|A_x, B_y, C_z)\! & = \!\sum_\xi p(\xi) P(b, c|B_y, C_z, \xi) P(a|A_x, \xi) \nonumber \\
&+ \!\sum_\xi q(\xi) P(a, c|A_x, C_z, \xi)P(b|B_y, \xi)\! \nonumber \\
&+\!\sum_\xi r(\xi) P(a, b|A_x, B_y, \xi)P(c|C_z, \xi), \label{HNLHV}
\end{align}
where $p(\xi)$, $q(\xi)$, $r(\xi)$ are three probability distributions over the hidden variables $\xi$; $\sum_\xi p(\xi)+\sum_\xi q(\xi)+\sum_\xi r(\xi)=1$.  The bipartite probability distribution $P(a, b|A_x, B_y, \xi)$ denotes the conditional probability of getting outcomes $a$ and $b$ when the measurement setting are $A_x$ and $B_y$, respectively, and $\xi$ is the hidden variable. $P(b, c|B_y, C_z, \xi)$ and $P(a, c|A_x, C_z, \xi)$ are defined similarly. Each of the bipartite probability distributions in the above decomposition can have arbitrary nonlocality.  

Genuine tripartite nonlocality is detected by violation of  the Svetlichny inequality \cite{SI}, which has the following form,
\begin{align}
S & = | \langle A_0 B_0 C_0 \rangle + \langle A_1 B_0 C_0 \rangle - \langle A_0 B_1 C_0 \rangle  + \langle A_1 B_1 C_0 \rangle \nonumber \\ 
&+ \langle A_0 B_0 C_1 \rangle - \langle A_1 B_0 C_1 \rangle + \langle A_0 B_1 C_1 \rangle  + \langle A_1 B_1 C_1 \rangle | \leq 4.
\label{si}
\end{align}
The magnitude of maximum quantum violation of Svetlichny inequality is $4\sqrt{2}$. GHZ state gives rise to this maximal quantum violation of Svetlichny inequality for a different choice of measurement settings which do not demonstrate GHZ paradox \cite{mermin2}.
 
\section{Setting up the scenario via weak measurement formalism} \label{scenario}
\begin{figure}[t!]
	\centering
	\includegraphics[scale=0.4]{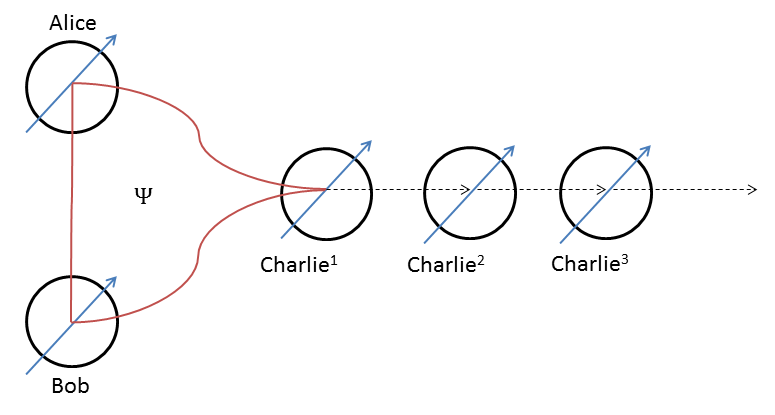}
	\caption{Sharing Tripartite Nonlocality Task: Consider a scenario where three spin-$\frac{1}{2}$ particles are prepared in the state $\psi$. These three particles are spatially separated and shared between Alice, Bob and multiple Charlies. Alice performs measurements on the first particle; Bob performs measurements on the second particle and multiple Charlies perform measurements on the third particle sequentially.}
	\label{fig1}
\end{figure}

Let us consider a scenario where three spin-$\frac{1}{2}$ particles are prepared in the state $\rho$. These three particles are spatially separated and shared between Alice, Bob and multiple Charlies (i. e., Charlie$^1$,  Charlie$^2$, Charlie$^3$, ..., Charlie$^n$). Alice performs measurements on the first particle; Bob performs measurements on the second particle and multiple Charlies perform measurements on the third particle sequentially. Initially, Charlie$^1$ performs measurements on the third particle and after doing measurements he delivers the particle to Charlie$^2$, Charlie$^2$ also passes the particle to Charlie$^3$ after doing measurements and so on. This scenario is depicted in Fig. \ref{fig1}. One important point to be stressed here is that, in this scenario considered by us, each Charlie performs measurements independent of the measurement choices and outcomes of the previous Charlies on the particle of his possession. Moreover, we are considering unbiased input scenario, i. e.,  all possible measurement settings of each Charlie are equally probable.

Note that in the above scenario no-signaling condition (the probability of obtaining one party's outcome does not depend on spatially separated other parties' settings) is satisfied between Alice, Bob and any Charlie as they are spatially separated and they perform measurements on three different particles. On the other hand, no-signaling condition is not satisfied between different Charlies. Because different Charlies perform measurements on the same particle sequentially. In fact, Charlie$^{m-1}$ signals to Charlie$^m$ (where $m$ $ \in \{2, ..., n\}$) by his choices of measurements on the state of the particle before he passes it on.

Against the above backdrop, we investigate how many Charlies can demonstrate tripartite nonlocality or genuine tripartite nonlocality with single Alice and single Bob. To be precise, we want to explore how many Charlies can have measurement statistics with single Alice and single Bob violating Mermin inequality (\ref{mi}) and Svetlichny inequality (\ref{si}), respectvely. In this case, each of the Charlies except the final Charlie cannot measure sharply. If any Charlie measures sharply, i.e., performs projective measurements, then there would be no possibility of violation of the Mermin inequality or Svetlichny inequality by the next Charlie, since the entanglement of the state shared between Alice, Bob and the subsequent Charlie would be completely destroyed. Hence, in order to address the aforementioned problem with $n$ Charlies, the measurements of the first $(n - 1)$ Charlies should be weak. In the present study we will follow the unsharp version of the weak measurement formalism discussed in \cite{mal,sas}. For completeness we briefly recapitulate in the following the weak measurement scheme introduced in \cite {sygp} and then unsharp version of that considered in \cite{mal,sas}.
  
In standard von Neuman measurement, after an interaction with a meter having the state $\phi (q)$,  the state $| \psi \rangle$ ($| \psi \rangle = a | 0 \rangle + b | 1 \rangle$, $|0\rangle$ and $|1\rangle$ form orthonormal basis in $\mathbb{C}^2$, $|a|^2+|b|^2 = 1$)  of a spin-$\frac{1}{2}$ particle becomes
\begin{equation}
a | 0 \rangle \otimes \phi (q-1)  + b | 1 \rangle  \otimes \phi (q+1).
\end{equation}
Weak version of this ideal measurement is characterised by two parameters: the quality factor $F$ and the precision $G$ of the measurements. Quality factor is given by, $F(\phi) = \int_{-\infty}^{\infty} \langle \phi (q+1) | \phi (q-1) \rangle dq$. It quantifies the extent to which the state of the system remains undisturbed after the measurement. Precision of measurement is defined as $ G = \int_{-1}^{1} \phi ^2 (q) dq$. It quantifies the information gain from measurement. In case of strong projective measurement, $F = 0$ and $G = 1$. An optimal pointer state is defined as the one which gives the best trade-off between these two quantities, i. e., for a given quality factor, it provides the greatest precision. It has been shown that the information-disturbance trade-off condition for an optimal pointer is given by, $F^2 + G^2 =1$ \cite{sygp}. 

This weak measurement formalism can be recast through the unsharp measurement formalism \cite{mal,sas}, which is a particular class of positive operator valued measurement (POVM) \cite{pb1,pb2}. POVM is a set of positive operators that add to identity, i. e.,  $E \equiv \{ E_i | \sum_i E_i = \mathbb{I}, 0 <E_i \leq \mathbb{I} \}$. Effects ($E_i$s) represent quantum events that may occur as outcomes of a measurement. In case of dichotomic unsharp measurement formalism, the effect operators are given by, 
\begin{equation}
E^\lambda_{\pm} = \lambda P_{\pm} + (1-\lambda) \frac{\mathbb{I}_2}{2},
\end{equation}
where $\lambda$ ($0 < \lambda \leq 1$) is the sharpness parameter, $P_{+}$ ($P_{-}$) is the projector associated with the outcome $+1$ ($-1$), $\mathbb{I}_2$ is the $2 \times 2$ identity matrix. The probability of getting the outcome $+1$ and $-1$ are $\text{Tr}[\rho E^\lambda_{+}]$ and $\text{Tr}[\rho E^\lambda_{-}]$, respectively, where $\rho$ is the state on which the measurement is performed. The post-measurement states are determined by Luder transformation rule. In case of above dichotomic unsharp measurement performed on the state $\rho$, the states after the measurements are given by $\dfrac{\sqrt{E^\lambda_{+}} \rho \sqrt{E^\lambda_{+}}}{\text{Tr}[E^\lambda_{+} \rho]}$ and $\dfrac{\sqrt{E^\lambda_{-}} \rho \sqrt{E^\lambda_{-}}}{\text{Tr}[E^\lambda_{-} \rho]}$ when $+1$ and $-1$ outcomes are obtained, respectively.

The quality factor $F$ and the precision $G$ defined in the context of aforementioned weak measurement formalism are related to the unsharp measurement formalism through $F = \sqrt{1-\lambda^2}$ and $G = \lambda$ \cite{mal,sas}. Hence, it is evident that the sharpness parameter $\lambda$ characterizes the precision of the measurement. For $G = \lambda = 1$, $F$ becomes zero, which is the case of sharp projective measurement. Hence, in unsharp measurement formalism, the optimal pointer state condition, $F^2 + G^2 = 1$, is automatically satisfied. In the following we will consider the measurements of all the Charlies except the final Charlie to be unsharp.

\section{Sharing of standard tripartite nonlocality by multiple Charlies} \label{sharetri}
In this Section we explore how many Charlies can simultaneously demonstrate standard tripartite nonlocality through the quantum violations of Mermin inequality (\ref{mi}) with single Alice and single Bob in the scenario discussed in Section \ref{scenario}. Note that Mermin inequality (\ref{mi}) is maximally violated by tripartite GHZ state \cite{GHZ} $\rho_{GHZ} = | \psi_{GHZ} \rangle \langle \psi_{GHZ} |$, where
\begin{equation}
|\psi_{GHZ} \rangle = \frac{1}{\sqrt{2}} ( |000 \rangle + | 111 \rangle ).
\label{ghz}
\end{equation}
Here $\{|0\rangle,|1\rangle\}$ form an orthonormal basis in $\mathbb{C}^2$. Hence, in order to probe optimal sharing of standard tripartite nonlocality through the quantum violation of Mermin inequality (\ref{mi}), we consider that Alice, Bob and multiple Charlies initially share tripartite GHZ state given by Eq. (\ref{ghz}). Suppose, Alice has a choice between two dichotomic measurements: spin component observables in the directions $\{ \hat{x}_0, \hat{x}_1 \}$ to perform; Bob has the choice between spin component observables in the directions $\{ \hat{y}_0, \hat{y}_1 \}$ to perform; and Charlie$^m$ (where $m$ $\in \{1, 2, ..., n\}$) has the choice between spin component observables in the directions $\{ \hat{z}_0^m, \hat{z}_1^m \}$ to perform. Outcomes of these measurements are labeled by $\{+1, -1 \}$.

We assume that the two possible choices of measurement settings of Alice are the spin component observables in the directions $\hat{x}_i$ given by,
\begin{equation}
\label{alicedir}
\hat{x}_i = \sin \theta^x_i \cos \phi^x_i \hat{X} + \sin \theta^x_i \sin \phi^x_i \hat{Y} + \cos \theta^x_i \hat{Z},
\end{equation}
where $0 \leq \theta^x_i  \leq \pi$; $0 \leq \phi^x_i  \leq 2 \pi$ and  $i\in\{0,1\}$. $\hat{X}$, $\hat{Y}$, $\hat{Z}$ are three orthogonal unit vectors in Cartesian coordinates. The two possible choices of measurement settings of Bob are the spin component observables in the directions $\hat{y}_i$ given by,
\begin{equation}
\label{bobdir}
\hat{y}_i = \sin \theta^{y}_i \cos \phi^{y}_i \hat{X} + \sin \theta^{y}_i \sin \phi^{y}_i \hat{Y} + \cos \theta^{y}_i \hat{Z},
\end{equation}
where $0 \leq \theta^y_i  \leq \pi$; $0 \leq \phi^y_i  \leq 2 \pi$ and  $i\in\{0,1\}$. Similarly, the two possible choices of measurement settings of Charlie$^m$ are the spin component observables in the directions $\hat{z_i^m}$ given by,
\begin{equation}
\label{charliemdir}
\hat{z}^m_i = sin \theta^{z^m}_i cos \phi^{z^m}_i \hat{X} + sin \theta^{z^m}_i sin \phi^{z^m}_i \hat{Y} + cos \theta^{z^m}_i \hat{Z},
\end{equation}
where $0 \leq \theta^{z^m}_i  \leq \pi$; $0 \leq \phi^{z^m}_i  \leq 2 \pi$ and  $i\in\{0,1\}$.

Suppose $P(a, b, c^1, c^2|\hat{x}_i, \hat{y}_j, \hat{z}^1_k, \hat{z}^2_l)$ denotes joint probability of obtaining the outcomes $a$, $b$, $c^1$, $c^2$ when Alice, Bob, Charlie$^1$ and Charlie$^2$ measures spin component observables in the directions $\hat{x}_i$, $\hat{y}_j$, $\hat{z}^1_k$ and $\hat{z}^2_l$, respectively ($a, b, c^1, c^2$ $\in \{ +1, -1\}$; $i , j, k, l$ $\in \{0, 1\}$). In the following we consider that Alice and Bob perform sharp measurements; Charlie$^m$ (where $m$ $\in \{1, 2, ..., (n-1)\}$) performs unsharp measurements with sharpness parameter $\lambda_m$. 

$P(a, b, c^1, c^2|\hat{x}_i, \hat{y}_j, \hat{z}^1_k, \hat{z}^2_l)$ can be evaluated using the Born rule given below:
\begin{equation}
P(a, b, c^1, c^2|\hat{x}_i, \hat{y}_j, \hat{z}^1_k, \hat{z}^2_l) = \text{Tr} \Big[ E^{\lambda_{2}}_{c^2} \cdot \rho^{C^2}_{un} \Big].
\label{j1}
\end{equation}
Here $E^{\lambda_{2}}_{c^2}$ = $\lambda_2 \dfrac{\mathbb{I}_2 + c^2 \hat{z}^2_l \cdot \vec{\sigma}}{2} + (1-\lambda_2) \dfrac{\mathbb{I}_2}{2}$; $\vec{\sigma}$ =  $(\sigma_1, \sigma_2, \sigma_3)$ is a vector composed of Pauli matrices.  $\rho^{C^2}_{un}$ is the unnormalized state at Charlie$^2$'s side when the outcomes $a$, $b$, $c^1$ are obtained by Alice, Bob, Charlie$^1$ by performing measurements of the spin component observables in the directions $\hat{x}_i$, $\hat{y}_j$, $\hat{z}^1_k$, respectively and $\rho_{un}^{C^2}$ is given by,
\begin{align}
\rho_{un}^{C^2} = \text{Tr}_{A B} \Bigg[ &\Big\{ \frac{\mathbb{I}_2+ a \hat{x}_i \cdot \vec{\sigma}}{2} \otimes \frac{\mathbb{I}_2 + b \hat{y}_j \cdot \vec{\sigma}}{2} \otimes \sqrt{E^{\lambda_{1}}_{c^1}} \Big\} \cdot \rho_{GHZ} \cdot \nonumber \\ 
&\Big\{ \frac{\mathbb{I}_2 + a \hat{x}_i \cdot \vec{\sigma}}{2} \otimes \frac{\mathbb{I}_2 + b \hat{y}_j \cdot \vec{\sigma}}{2} \otimes \sqrt{E^{\lambda_{1}}_{c^1}} \Big\} \Bigg],
\end{align}
where,
\begin{equation}
\sqrt{E^{\lambda_{1}}_{c^1}} = \sqrt{\dfrac{1+\lambda_1}{2}} \dfrac{\mathbb{I}_2 + c^1 \hat{z}^1_k \cdot \vec{\sigma}}{2} +\sqrt{\dfrac{1- \lambda_1}{2}} \dfrac{\mathbb{I}_2 - c^1 \hat{z}^1_k \cdot \vec{\sigma}}{2}
\end{equation}

$\text{Tr}_{AB}[...]$ denotes partial trace over the subsystems of Alice and Bob. From Eq.(\ref{j1}) one can obtain $P(a, b, c^2|\hat{x}_i, \hat{y}_j, \hat{z}^1_k, \hat{z}^2_l)$, the joint probability of obtaining the outcomes $a$, $b$, $c^2$ when Alice, Bob, Charlie$^2$ measures spin component observables in the directions $\hat{x}_i$, $\hat{y}_j$, $\hat{z}^2_l$, respectively and given that Charlie$^1$ has performed spin component observables in the directions $\hat{z}^1_k$,:
\begin{equation}
P(a, b, c^2|\hat{x}_i, \hat{y}_j, \hat{z}^1_k, \hat{z}^2_l) = \sum_{c^1 = +1, -1} P(a, b, c^1, c^2|\hat{x}_i, \hat{y}_j, \hat{z}^1_k, \hat{z}^2_l).
\label{j2}
\end{equation}

Let $C_{ijkl}^2$ denotes the correlation between Alice, Bob and Charlie$^2$ when Alice, Bob, Charlie$^1$ and Charlie$^2$ measures spin component observables in the directions $\hat{x}_i$, $\hat{y}_j$, $\hat{z}^1_k$ and $\hat{z}^2_l$, respectively. From the above joint probability (\ref{j2}) the correlation $C_{ijkl}^2$ can be obtained in the following way,
\begin{equation}
C_{ijkl}^2 = \sum_{a = +1, -1} \sum_{b = +1, -1} \sum_{c^2 = +1, -1} (a b c^2) P(a, b, c^2|\hat{x}_i, \hat{y}_j, \hat{z}^1_k, \hat{z}^2_l).
\label{corr}
\end{equation}

Since Charlie$^2$ is ignorant about the measurement settings of Charlie$^1$, the above correlation has to be averaged over the two possible measurement settings of Charlie$^1$ (spin component observables in the directions $\{ \hat{z}^1_0, \hat{z}^1_1 \}$). This average correlation function between Alice, Bob and Charlie$^2$ is given by,
\begin{equation}
\overline{C_{ijl}^2} = \sum_{k = 0,1} C_{ijkl}^2 P(\hat{z}^1_k),
\label{avcorr}
\end{equation}
where $P(\hat{z}^1_k)$ is the probability with which Charlie$^1$ performs measurement of spin component observables in the direction $\hat{z}^1_k$ ($k \in \{ 0, 1 \}$). Since we are considering an unbiased input scenario, all the possible measurement settings of Charlie$^1$ are equally probable, i. e.,  $P(\hat{z}^1_0)$ = $P(\hat{z}^1_1)$ = $\frac{1}{2}$.

In a similar way the average correlation between Alice, Bob and any Charlie$^m$, $\overline{C_{ijl}^m}$ ($i, j, l$ $\in$ $\{0, 1\}$, $m$ $\in$ $\{1, 2, 3, ..., n \}$), can be evaluated. Using these average correlations the Mermin inequality (\ref{mi}) for Alice, Bob and Charlie$^m$ can be expressed as follows:
\begin{equation}
M_m = |\overline{C_{100}^m} +  \overline{C_{010}^m} + \overline{C_{001}^m}   - \overline{C_{111}^m} | \leq 2,
\label{mim}
\end{equation}
whose violation implies that standard tripartite nonlocality is demonstrated by Alice, Bob and Charlie$^m$.

Now we want to find out whether Charlie$^1$ and Charlie$^2$ can  simultaneously demonstrate standard tripartite nonlocality with Alice and Bob. Consider the measurements of the final Charlie, i. e., Charlie$^2$ to be sharp ($\lambda_2 = 1$), and the measurements of Charlie$^1$ to be unsharp. We observe that, when Charlie$^1$ gets $5\%$ violation of the Mermin inequality (\ref{mim}), i. e., when $M_1 = 2.10$, then the maximum quantum violation of Mermin inequality (\ref{mim}) obtained by Charlie$^2$ is $85 \%$, i.e., $M_2 = 3.70$. This happens for the choice of measurement settings: ($\theta^x_0$, $\phi^x_0$, $\theta^x_1$, $\phi^x_1$, $\theta^y_0$, $\phi^y_0$, $\theta^y_1$, $\phi^y_1$, $\theta^{z^1}_0$, $\phi^{z^1}_0$, $\theta^{z^1}_1$, $\phi^{z^1}_1$,  $\theta^{z^2}_0$, $\phi^{z^2}_0$, $\theta^{z^2}_1$, $\phi^{z^2}_1$) $\equiv$ ($\frac{\pi}{2}$, $\frac{\pi}{2}$, $\frac{\pi}{2}$, $0$, $\frac{\pi}{2}$, $\frac{\pi}{2}$, $\frac{\pi}{2}$, $0$, $\frac{\pi}{2}$, $\frac{\pi}{2}$, $\frac{\pi}{2}$, $0$, $\frac{\pi}{2}$, $\frac{\pi}{2}$, $\frac{\pi}{2}$, $0$)  with $\lambda_1 =0.52$. 

Next, we address the question whether Charlie$^1$, Charlie$^2$ and Charlie$^3$ can  simultaneously demonstrate standard tripartite nonlocality with Alice and Bob. In this case, the measurements of the final Charlie, i. e., Charlie$^3$ are sharp ($\lambda_3 = 1$), and the measurements of Charlie$^1$ and Charlie$^2$ are unsharp. In this case we observe that, when each of the quantum violations of Mermin inequality (\ref{mim}) by Alice, Bob, Charlie$^1$ and Alice, Bob, Charlie$^2$ is $5\%$, i. e., when $M_1 = 2.10$ and $M_2 = 2.10$, then the maximum quantum violation of Mermin inequality (\ref{mim}) by Alice, Bob, Charlie$^3$ is $69 \%$, i. e., $M_3 = 3.38$. This happens for the choice of measurement settings: ($\theta^x_0$, $\phi^x_0$, $\theta^x_1$, $\phi^x_1$, $\theta^y_0$, $\phi^y_0$, $\theta^y_1$, $\phi^y_1$, $\theta^{z^1}_0$, $\phi^{z^1}_0$, $\theta^{z^1}_1$, $\phi^{z^1}_1$,  $\theta^{z^2}_0$, $\phi^{z^2}_0$, $\theta^{z^2}_1$, $\phi^{z^2}_1$, $\theta^{z^3}_0$, $\phi^{z^3}_0$, $\theta^{z^3}_1$, $\phi^{z^3}_1$) $\equiv$ ($\frac{\pi}{2}$, $\frac{\pi}{2}$, $\frac{\pi}{2}$, $0$, $\frac{\pi}{2}$, $\frac{\pi}{2}$, $\frac{\pi}{2}$, $0$, $\frac{\pi}{2}$, $\frac{\pi}{2}$, $\frac{\pi}{2}$, $0$, $\frac{\pi}{2}$, $\frac{\pi}{2}$, $\frac{\pi}{2}$, $0$, $\frac{\pi}{2}$, $\frac{\pi}{2}$, $\frac{\pi}{2}$, $0$)  with $\lambda_1=0.52$ and $\lambda_2=0.57$. Hence, it is possible for Charlie$^1$, Charlie$^2$ and Charlie$^3$ to  simultaneously demonstrate standard tripartite nonlocality with Alice and Bob.

Proceeding in a similar way we observe that Charlie$^1$, Charlie$^2$, Charlie$^3$, Charlie$^4$, Charlie$^5$, Charlie$^6$ can  simultaneously demonstrate standard tripartite nonlocality with Alice and Bob. Now we want to investigate whether Charlie$^1$, Charlie$^2$, Charlie$^3$, Charlie$^4$, Charlie$^5$, Charlie$^6$ and Charlie$^7$ can  simultaneously demonstrate standard tripartite nonlocality with single Alice and single Bob. Here the measurements of the final Charlie, i. e., Charlie$^7$ are sharp ($\lambda_7 = 1$), and the measurements of all other Charlies are unsharp. In this case we observe that, when each of the quantum violations of Mermin inequality (\ref{mim}) by Charlie$^1$, Charlie$^2$, Charlie$^3$, Charlie$^4$, Charlie$^5$, Charlie$^6$ is $2.5\%$, i. e., when $M_1 = 2.05$, $M_2 = 2.05$, $M_3 = 2.05$, $M_4 = 2.05$, $M_5 = 2.05$, $M_6 = 2.05$, then the maximum quantum value of the left hand side of Mermin inequality (\ref{mim}) by Alice, Bob, Charlie$^7$ is $M_7 = 1.49$. In fact, it is observed that when $M_1 = 2$, $M_2 = 2$, $M_3 = 2$, $M_4 = 2$, $M_5 = 2$, $M_6 = 2$, then the maximum quantum value of the left hand side of Mermin inequality (\ref{mim}) by Alice, Bob, Charlie$^7$ is $M_7 = 1.76$. Hence, Charlie$^1$, Charlie$^2$, Charlie$^3$, Charlie$^4$, Charlie$^5$, Charlie$^6$, Charlie$^7$ cannot demonstrate standard tripartite nonlocality via violation of Mermin inequality with Alice and Bob simultaneously.

It is to be noted here that Charlie$^7$ may obtain quantum mechanical violation of the Mermin inequality if the sharpness parameter of any previous Charlie is small enough not to get a violation. In fact, it can be easily checked that Alice and Bob can demonstrate standard tripartite nonlocality through the quantum violation of Mermin inequality (\ref{mim}) with at most any six Charlies on one side, simultaneously.

Though Mermin inequality is not maximally violated by W-state, given by $|\psi_W \rangle$ = $\frac{1}{\sqrt{3}} (|001 \rangle + |010 \rangle + |100 \rangle)$, we also investigate the above issue when W state is initially shared between Alice, Bob and multiple Charlies as W state and GHZ state are two inequivalent three qubit maximally entangled states \cite{two}. It is observed that at most three Charlies can demonstrate standard tripartite nonlocality through the quantum violation of Mermin inequality (\ref{mi}) with single Alice and single Bob in the scenario described in Section \ref{scenario} when W state is shared.

In the next Section we address the same question discussed above in the context of genuine tripartite nonlocality.
 
\section{Sharing of genuine tripartite nonlocality by multiple Charlies}\label{sharetri2}

Since tripartite nonlocality and genuine tripartite nonlocality are fundamentally different, we also investigate how many Charlies can demonstrate genuine tripartite nonlocality through the quantum violations of Svetlichny inequality (\ref{si}) with single Alice and single Bob in the scenario discussed in Section \ref{scenario}. Since Svetlichny inequality (\ref{si}) is maximally violated by tripartite GHZ state (\ref{ghz}), in this case also we consider that GHZ state is initially shared between Alice, Bob and multiple Charlies.  The two possible choices of measurement settings of Alice, Bob and Charlie$^m$ are the spin component observables in the directions $\hat{x}_i$ given by Eq.(\ref{alicedir}), $\hat{y}_i$ given by Eq.(\ref{bobdir}) and $\hat{z}^m_i$ given by Eq.(\ref{charliemdir}), respectively, where $i\in\{0,1\}$. Outcomes of these measurements are labeled by $\{-1, +1 \}$.

Using the average correlations $\overline{C_{ijl}^m}$ between Alice, Bob and Charlie$^m$ calculated by following the procedure described in Section \ref{sharetri}, the Svetlichny inequality (\ref{si}) for Alice, Bob and Charlie$^m$ can be expressed as follows:
\begin{align}
S_m  = & | \overline{C_{000}^m} +\overline{C_{100}^m}  - \overline{C_{010}^m}  + \overline{C_{110}^m} + \overline{C_{001}^m} - \overline{C_{101}^m} + \overline{C_{011}^m}  + \overline{C_{111}^m} | \nonumber \\ 
&\leq 4,
\label{sis}
\end{align}
whose violation implies that genuine tripartite nonlocality is demonstrated by Alice, Bob and Charlie$^m$.

Now we are going to find out whether Charlie$^1$ and Charlie$^2$ can  simultaneously demonstrate genuine tripartite nonlocality with Alice and Bob. Here the measurements of the final Charlie, i. e., Charlie$^2$ are sharp ($\lambda_2 = 1$), and the measurements of Charlie$^1$ are unsharp. We observe that, when Charlie$^1$ gets $5\%$ violation of the Svetlichny inequality (\ref{sis}), i. e., when $S_1 = 4.20$, then the maximum quantum violation of Svetlichny inequality (\ref{sis}) obtained by Charlie$^2$ is $18\%$, i.e., $S_2 = 4.72$. This happens for the choice of measurement settings: ($\theta^x_0$, $\phi^x_0$, $\theta^x_1$, $\phi^x_1$, $\theta^y_0$, $\phi^y_0$, $\theta^y_1$, $\phi^y_1$, $\theta^{z^1}_0$, $\phi^{z^1}_0$, $\theta^{z^1}_1$, $\phi^{z^1}_1$,  $\theta^{z^2}_0$, $\phi^{z^2}_0$, $\theta^{z^2}_1$, $\phi^{z^2}_1$) $\equiv$ ($\frac{\pi}{2}$, $\frac{\pi}{2}$, $\frac{\pi}{2}$, $0$, $\frac{\pi}{2}$, $\frac{\pi}{2}$, $\frac{\pi}{2}$, $0$, $\frac{\pi}{2}$, $\frac{\pi}{4}$, $\frac{\pi}{2}$, $\frac{3\pi}{4}$, $\frac{\pi}{2}$, $\frac{\pi}{4}$, $\frac{\pi}{2}$, $\frac{3\pi}{4}$)  with $\lambda_1 =0.74$. Most importantly, it is observed that for $\lambda_1 \in [0.71, 0.91]$ both Charlie$^1$ and Charlie$^2$ can demonstrate genuine tripartite nonlocality with Alice and Bob simultaneously.

Next, we are interested in the question whether Charlie$^1$, Charlie$^2$ and Charlie$^3$ can demonstrate genuine tripartite nonlocality with single Alice and single Bob simultaneously. In this case, the measurements of the final Charlie, i. e., Charlie$^3$ are sharp ($\lambda_3 = 1$), and the measurements of Charlie$^1$ and Charlie$^2$ are unsharp. In this case we observe that, when each of the quantum violations of Svetlichny inequality (\ref{sis}) by Alice, Bob, Charlie$^1$ and Alice, Bob, Charlie$^2$ is $5\%$, i. e., when $S_1 = 4.20$ and $M_2 = 4.20$, then the maximum quantum value of the left hand side of Svetlichny inequality (\ref{sis}) by Alice, Bob, Charlie$^3$ is $S_3 = 3.44$. In fact, when $S_1 = 2$, $S_2 = 2$, then the maximum quantum value of the left hand side of Svetlichny inequality (\ref{sis}) by Alice, Bob, Charlie$^3$ is $S_3 = 3.77$. This happens for the choice of measurement settings: ($\theta^x_0$, $\phi^x_0$, $\theta^x_1$, $\phi^x_1$, $\theta^y_0$, $\phi^y_0$, $\theta^y_1$, $\phi^y_1$, $\theta^{z^1}_0$, $\phi^{z^1}_0$, $\theta^{z^1}_1$, $\phi^{z^1}_1$,  $\theta^{z^2}_0$, $\phi^{z^2}_0$, $\theta^{z^2}_1$, $\phi^{z^2}_1$, $\theta^{z^3}_0$, $\phi^{z^3}_0$, $\theta^{z^3}_1$, $\phi^{z^3}_1$) $\equiv$ ($\frac{\pi}{2}$, $\frac{\pi}{2}$, $\frac{\pi}{2}$, $0$, $\frac{\pi}{2}$, $\frac{\pi}{2}$, $\frac{\pi}{2}$, $0$, $\frac{\pi}{2}$, $\frac{\pi}{4}$, $\frac{\pi}{2}$, $\frac{3\pi}{4}$, $\frac{\pi}{2}$, $\frac{\pi}{4}$, $\frac{\pi}{2}$, $\frac{3\pi}{4}$, $\frac{\pi}{2}$, $\frac{\pi}{4}$, $\frac{\pi}{2}$, $\frac{3\pi}{4}$)  with $\lambda_1=0.71$ and $\lambda_2=0.83$. Hence, it is impossible for Charlie$^1$, Charlie$^2$ and Charlie$^3$ to  simultaneously demonstrate genuine tripartite nonlocality via violation of Svetlichny inequality with Alice and Bob.

It is to be noted here that Charlie$^3$ may obtain quantum mechanical violation of the Svetlichny inequality if the sharpness parameter of Charlie$^1$ or that of Charlie$^2$ is small enough not to get a violation. In fact, it can be easily checked that Alice and Bob can demonstrate genuine tripartite nonlocality through the quantum violation of Svetlichny inequality (\ref{sis}) with any of the combinations: (Charlie$^1$, Charlie$^2$), (Charlie$^1$, Charlie$^3$), (Charlie$^2$, Charlie$^3$), simultaneously.

In this case also we investigate the above issue when W state is initially shared between Alice, Bob and multiple Charlies. It is observed that at most one Charlie can demonstrate genuine tripartite nonlocality through the quantum violation of Svetlichny inequality (\ref{si}) with single Alice and single Bob in the scenario described in Section \ref{scenario} when W state is shared.

\section{Concluding discussion}\label{conclusion}
Tripartite as well as multipartite quantum correlations have various applications in quantum communication and quantum computation \cite{ap1,ap2,ap3,ap4,ap5,ap6,ap7,ap8,ap9,ap10,ap11,ap12,ap13,ap14}. Hence, from foundational aspect as well as from quantum information theoretic perspective it is legitimate to ask whether more than three observers can share tripartite nonlocality, simultaneously. In the present study, contingent upon considering a particular scenario, we have answered this question in the affirmative. Moreover, we have demonstrated that standard tripartite nonlocality can be simultaneously shared by more number of observers (Charlies), performing sequential measurements on one particle, compared to genuine tripartite nonlocality in the particular scenario considered in Fig.\ref{fig1}. Hence, we can state that standard tripartite nonlocality is more effective than genuine tripartite nonlocality in the context of simultaneous sharing by multiple observers.

In particular, we have considered a scenario where three spin-$\frac{1}{2}$ particles are spatially separated and shared between Alice, Bob and multiple Charlies. Alice performs measurements on the first particle; Bob performs measurements on the second particle and multiple Charlies perform measurements on the third particle sequentially. Moreover, we have assumed that all possible measurement settings of each Charlie are equally probable and each Charlie performs measurements independent of the choices of measurement settings and outcomes of previous Charlies. In this scenario we have investigated how many Charlies can simultaneously show tripartite nonlocality with single Alice and single Bob. Interestingly, we have shown that at most six Charlies can simultaneously demonstrate standard tripartite nonlocality with respect to quantum violations of Mermin inequality \cite{mermin}. On the other hand, at most two Charlies can simultaneously demonstrate genuine tripartite nonlocality with respect to quantum violations of Svetlichny inequality \cite{SI}. Apart from having information theoretic applications, these results distinguish two inequivalent notions of tripartite nonlocality in the context of nonlocality sharing.

The present work opens a number of interesting questions. Firstly, addressing the issue of sharing other notions of genuine tripartite  nonlocality \cite{B}, apart from Svetlichny type genuine nonlocality, in the aforementioned scenario described in Fig.\ref{fig1} is worthy of further investigation. Secondly, in the same spirit of the present work, it would be interesting to investigate sharing of multipartite quantum steering \cite{stm1,stm2,stm3} by multiple observers measuring sequentially on the same particle. Thirdly, exploring the concept of nonlocality sharing for higher dimensional quantum systems by using different many-outcome \emph{local realist} inequalities \cite{cglmp,gwi} remains open till date. Another interesting direction is to explore the possibility of sharing tripartite nonlocal correlations between (i) single Alice - multiple Bobs - multiple Charlies, (ii)  multiple Alices - multiple Bobs - multiple Charlies and finding a connection of these sharing phenomena with some information processing task(s). It is to be noted that the problem presented in this paper can easily be generalized to multipartite nonlocality, demonstrated by multiqubit states, using different multipartite Bell-type \emph{local realist} inequalities \cite{mermin,multi2,multi3,multi4,multi5,multi6}.

\section{ACKNOWLEDGEMENTS}
D. D. acknowledges the financial support from University Grants Commission (UGC), Government of India. S. S. acknowledges the financial support from INSPIRE programme, Department of Science and Technology (DST), Government of India. The authors acknowledge fruitful discussions with Biswajit Paul.

\end{document}